\documentclass{aa}
\usepackage{graphicx}
\usepackage{amsmath}

\begin{document}
\thesaurus{2(12.07.1; 02.07.1; 11.17.3; 02.23.1; 03.13.4)}

\title{Quasar luminosity and twin effects induced by filamentary and planar
structures}
\author{Salvatore Capozziello\thanks{%
E-mail capozziello@sa.infn.it} and Gerardo Iovane\thanks{%
E-mail geriov@sa.infn.it}} \institute{Dipartimento di
Scienze Fisiche E.R. Caianiello,\\
 Universit\`{a} di Salerno, I-84081 Baronissi, Salerno, Italy.\\
 Istituto Nazionale di Fisica Nucleare, sezione di Napoli.}
\date{Received / Accepted }
\authorrunning{S. Capozziello \& G. Iovane}
\maketitle

\begin{abstract}
We consider filamentary and planar large scale structures as
possible refraction channels for electromagnetic radiation coming
from cosmological structures. By this hypothesis, it is possible
to explain the quasar luminosity distribution and, in particular,
the presence of "twin" and "brother" objects. Method and details
of simulation are given.
\end{abstract}

\section{Introduction}

Gravitational lensing (Straumann et al. 1998) is today considered
a basic tool to investigate the large scale structure of the
universe and to perform tests of cosmological models. Furthermore,
it plays a relevant role in the dark matter search from small to
large scales. On the other hand, the lensing effects can be
explained by considering the action of a weak gravitational field
deflecting the light rays which come from a fixed source. In other
words, it is possible to define a refraction index, connected to
the Newtonian gravitational field.

However, it is worthwhile to note that, in optics, there are other
kinds of effects and instruments, like optical fibres and
waveguides, which use the same deflection principles: The analogy
with the gravitational field can be extended also to these
processes taking into account a sort of ''nonlinearity'' in
gravitational lensing.

Considering the concept of ''channel'' of sister sciences, we have
refraction channels, acoustic channels and several other examples where
propagating waves are trapped by some structures.

In cosmology, matter distributions, able to produce gravitational
potential inducing such ''channelling effects'', could explain
several anomalies which are hard to connect to standard
gravitational lensing models, like, for example, the existence of
objects with the same spectrum and redshift, but placed in the sky
at large angular distances (e.g. twins). The huge luminosity of
quasars (Fan et al. 2000; Zheng et al. 2000) could be explained by
using planar cosmological structures too. In fact, taking into
account expanding bubbles, produced during first order phase
transition, a matter density gradient is generated when they come
in touch each other. Such expanding bubbles create tridimensional
spatial structures like hexahedrons (e.g. honeycombs) to minimize
the interstitial space. However the growth is different with
respect to crystal models since in these last cases the growth is
stopped in such a direction as well as a cell of the lattice
touches with an other expanding crystal in the same direction. In
cosmology, we have to take into account the expansion of the
universe.

Our hypothesis is that the surfaces of such hexahedrons, which are contact
surfaces, could be approximated by planar structures acting as channels for
electromagnetic radiation.

Furthermore, as we will discuss below, also filamentary structures could act
like channels too.

In principle, all primordial topological defects as cosmic
strings, textures and domain walls (Vilenkin 1984; Vilenkin 1985)
could be evolved so that today they can trap electromagnetic or
gravitational radiation (Capozziello \& Iovane 1998; Turner et al.
1986).

This gravitational waveguiding effect has the same physical reason of the
gravitational lensing effect, which account for the light deflection given
by a gravitational field acting like a refraction medium . However, there
are some substantial differences: the gravitational lenses are usually
compact objects. On the contrary, the density can be low in these
gravitational channels; moreover the radiation is deflected by the lens, but
it does not pass through the object, while the light have to travel
''inside'' the matter distribution in the case of gravitational channels.

The existence of a waveguiding effect could explain the huge luminosity of
very far objects like quasars. Another effect could be the generation of
twins and brothers. In other words, an observer could detect two images of
the same object: The real image and the one catched by gravitational channel.

From a theoretical point of view, the matter distribution should
locally produce an effective gravitational potential of the form
$\Phi \left( r\right) \sim r^{2}$.

It is worthwhile to note that the specific form of $r$ (and then of the
effective gravitational potential) depends on the particular symmetry of the
situation. As we shall discuss below, we have $r=x$ for ''planar'' mass
distributions while $r=\sqrt{x^{2}+y^{2}}$ in the case of ''filamentary''
distributions. In both cases, we are assuming an almost constant density $%
\rho $ which is substantially different with respect to the background.

The planar or filamentary distributions, essentially connected to a gradient
of matter density, could be generated at the contact surfaces among several
expanding bubbles.

The paper is organized as follows. In the Sec.2, the theoretical
model for gravitational channels is developed. Sec.3 is devoted to
the discussion of early cosmological phase transition, topological
defects and possible connections to the today observed large scale
structures. In Sec. 4, we present a simulation by which we try to
explain the luminosity distribution of quasars assuming \ that a
part of them has been ''twined'' because the radiation emitted
from them have undergone a channelling process by intervening
filamentary or planar large scale structures. The results are
discussed in Sec.5. Conclusions are drawn in Sec.6.

\section{Light propagation by gravitational refraction channels}

The behaviour of the electromagnetic field, without source and in
the presence of a gravitational field (Schneider, Ehlers, Falco
1992) can be described by the Maxwell equations
\begin{equation}
\frac{\partial F_{\alpha \beta }}{\partial x^{\gamma }}+\frac{\partial
F_{\beta \gamma }}{\partial x^{\alpha }}+\frac{\partial F_{\gamma \alpha }}{%
\partial x^{\beta }}=0;\quad \frac{1}{\sqrt{-g}}\frac{\partial }{\partial
x^{\beta }}\left( \sqrt{-g}F^{\alpha \beta }\right) =0,
\end{equation}

where $F^{\alpha \beta }$ is the electromagnetic field tensor and $\sqrt{-g}$
is the determinant of the four--dimensional metric tensor. For a static
gravitational field, these equations can be reduced to the usual Maxwell
equations describing the electromagnetic field in media where the dielectric
and magnetic tensor permeabilities are connected to the metric tensor $%
g_{\mu \nu }$ by the equation

\begin{equation}
\varepsilon _{ik}=\mu _{ik}=-g_{00}^{-1/2}[\mbox{det}g_{ik}]^{-1/2}g_{ik}%
\quad i,k=1,2,3.
\end{equation}

If \ the medium is isotropic, the metric tensor is diagonal and the
refraction index is $n\left( r\right) =\left( \varepsilon \mu \right) ^{%
\frac{1}{2}}$ (Capozziello \& Iovane 1999).

In the weak field approximation, the metric tensor components are
(Blioh \& Minakov 1989; Landau \& Lifsits 1975)

\begin{equation}
g_{00}\simeq 1+2\frac{\Phi (\mathbf{r})}{c^{2}}\;;\qquad g_{ik}\simeq
-\delta _{ik}\left( 1-2\frac{\Phi (\mathbf{r})}{c^{2}}\right) \;,
\end{equation}

where, $\Phi (\mathbf{r})$ is the Newtonian \ potential and we are
assuming the weak field, $\Phi /c^{2}\ll 1$, and the slow motion
approximation $|v|\ll c $.

Channeling solutions can be obtained, by reducing Maxwell
equations in a
medium to the scalar Helmoltz equation for the fields $\mathbf{E}$ and $%
\mathbf{H}$ . For the electrical field, after Fourier
transformation, we can write (Capozziello \& Iovane 1998)

\begin{equation}
\triangle \mbox{\bf E}_{\omega }(\mbox{\bf r})+\frac{\omega ^{2}}{c^{2}}%
n^{2}(\omega ,\mbox{\bf r})\mbox{\bf E}_{\omega }(\mbox{\bf r})=0\,
\end{equation}

where $\omega $ is the frequency of the given Fourier component.

A similar equation holds for the magnetic field. The general solution has
the form

\begin{equation}
\mbox{\bf E}_{\omega }(\mbox{\bf r})=\mbox{\bf E}_{\omega }^{(0)}(%
\mbox{\bf
r})+\int \mathcal{G}({\bf \tilde{r}},{\bf r}^{\prime },\omega )\nabla %
\left[ \frac{{\bf E}_{\omega }^{(0)}({\bf \tilde{r}})\nabla
\varepsilon
(\omega ,{\bf \tilde{r}})}{\varepsilon (\omega ,{\bf \tilde{r}})}\right] d%
{\bf \tilde{r}},
\end{equation}

where $\mathcal{G}$ is the Green function for the equation

\begin{equation}
\left[ \triangle +\frac{\omega ^{2}}{c^{2}}n^{2}(\omega ,\mbox{\bf r})\right]
\mathcal{G}(\mbox{\bf r},\mbox{\bf r}^{\prime },\omega )=\delta (\mbox{\bf r}-%
\mbox{\bf r}^{\prime })\,.
\end{equation}

This is a Schr\"{o}dinger-like equation for the energy constant $E=0$. In
fact, if we write down the Hamiltonian operator $\hat{\mathcal{H}}=-\frac{1}{%
2}\triangle +U(\mbox{\bf r})\,,$with $\hbar =m=1$ and ${\displaystyle U({\bf r})=-2%
\frac{\omega ^{2}}{c^{2}}n^{2}(\omega ,{\bf r})\,},$ a fully
analogy between the two problems is obtained. $U(\mbox{\bf r})$\
assumes the role of the Newtonian potential $\Phi (\mathbf{r})$.

Let us consider the scalar Helmoltz equation for an arbitrary monochromatic
component of the electric field

\begin{equation}
\frac{\partial ^{2}E}{\partial z^{2}}+\frac{\partial ^{2}E}{\partial x^{2}}+%
\frac{\partial ^{2}E}{\partial y^{2}}+k^{2}n^{2}(\mathbf{r})E=0\,,
\end{equation}

where $k$ is the wavenumber. The coordinate $z$ can be considered as the
longitudinal one and it can measure the space distance along the structure
produced by a mass distribution with an optical axis.

Let us consider now a solution of the form

\begin{equation}
E=n_{0}^{-1/2}\Psi \exp \left( ik\int^{z}n_{0}(z^{\prime })dz^{\prime
}\right) \;;\quad n_{0}\equiv n(0,0,z)\,,
\end{equation}

where $\Psi (x,y,z)$ is a slowly varying spatial amplitude along
the $z$ axis, and $\exp(iknz)$ is a rapidly oscillating phase
factor. Its clear that the beam propagation is along the $z$ axis.
We rewrite the Helmoltz equation neglecting the second order
derivative in longitudinal coordinate $z$ and obtain a
Schr\"{o}dinger--like equation for $\Psi $:

\begin{eqnarray}
& i\lambda \frac{\partial \Psi }{\partial \xi }= &\nonumber \\
&=-\frac{\lambda ^{2}}{2}\left( \frac{\partial ^{2}\Psi }{\partial
x^{2}}+\frac{\partial ^{2}\Psi }{\partial y^{2}}\right)
+\frac{1}{2}\left[ n_{0}^{2}(z)-n^{2}(x,y,z)\right] \Psi ,
\end{eqnarray}

where $\lambda $ is the electromagnetic radiation wavelength and
we adopt the new variable ${\displaystyle\xi
=\int^{z}\frac{dz^{\prime }}{n_{0}(z^{\prime })}\,}$ is a new
normalized variable with respect to the refraction index. It is
interesting to stress that the role of Planck constant is assumed by $%
\lambda $.

At this point, it is worthwhile to note that if one has the
distribution of matter in the form of a cylinder or a sphere with
a constant (dust) density, the gravitational potential inside has
a parabolic profile providing channelling effects for
electromagnetic radiation analogous to the sel-foc optical
waveguides realized for optical fibres. In this case,
Schr\"{o}dinger-like equation is a two-dimensional quantum
harmonic oscillator which solutions provs the form of
Gauss-Hermite polynomials (see, for example (Man'ko 1986)).

The channeling effect depends explicitly on the shape of the
potential: The radiation from a remote source does not attenuate
if $U\sim r^{2}$; this situation is realized supposing a
''planar'' mass distribution ($r=x$) or a filamentary distribution
($r=\sqrt{x^{2}+y^{2}}$) with constant density . In other words,
if the radiation, travelling from some source, undergoes a
channelling effect, it does not attenuate like $1/R^{2}$ as usual,
but it is, in some sense conserved; this fact means that the
source brightness will turn out to be much stronger than the
brightness of analogous objects located at the same distance (i.e.
at the same redshift Z) and the apparent energy released by the
source will be anomalously large.

To fix the ideas, let us estimate how the electric field propagates into an
ideal filament whose internal potential is

\begin{equation}
U(r)=\frac{1}{2}\omega ^{2}r^{2}\,,\;\;\;\;\;\;\omega ^{2}=\frac{4\pi G\rho
}{c^{2}}.
\end{equation}

A spherical wave from a source has the form

\begin{equation}
E=(1/R)\exp (ikR)\,.
\end{equation}

In the paraxial approximation, Eq. (11) becomes

\begin{equation}
E(z,r)=\frac{1}{z}\exp \left( ikz+\frac{ikr^{2}}{2z}-\frac{r^{2}}{2z^{2}}%
\right) \,,
\end{equation}

where we are using the expansion

\begin{equation}
R=\left( z^{2}+r^{2}\right) ^{1/2}\approx z\left( 1+\frac{r^{2}}{2z^{2}}%
\right) ,\;\;r\ll z\,.
\end{equation}

It is realistic to assume $n_{0}\simeq 1$ so that $\xi =z$.

Let us consider now that the starting point of the gravitational
channel of length $L$ is at a distance $l$ from a source shifted
by a distance $a$ from the structure longitudinal axis in the x
direction. The amplitude of the field $E$, entering the channel,
is

\begin{equation}
\Psi _{in}=\frac{1}{l}\exp \left[ \frac{ikl-1}{2l^{2}}\left(
(x-a)^{2}+y^{2}\right) \right] \,,
\end{equation}

so that we have $R=\left( l^{2}+y^{2}+(x-a)^{2}\right) ^{1/2}.$ We can
calculate the amplitude of the field at the exit of the filament by the
equation

\begin{eqnarray}
&\Psi _{out} &(x,y,l+L)= \nonumber \\
&& =\int dx_{1}dy_{1}\mathcal{G}(x,y,l+L,x_{1},y_{1},l)%
\Psi _{in}(x_{1},y_{1},l),
\end{eqnarray}

where $\mathcal{G}$ is again the Green function for $\Psi $. In particular,
taking into account the potential (10), $\mathcal{G}$ is the propagator of
the harmonic oscillator; then $\Psi _{out}$ is a Gaussian integral that can
be exactly evaluated. There are two interesting limit for $\Psi _{out}$.

If $\omega l\ll 1$, we have

\begin{equation}
\Psi _{out}=\frac{1}{i\lambda }\exp \left\{ -\frac{l+i\lambda }{2\lambda
^{2}l}\left[ (x+a)^{2}+y^{2}\right] \right\} \,,
\end{equation}

which means that the radiation emerging from a point with coordinate $%
(a,0,0) $ is focused near a point with coordinates $(-a,0,l+L)$ (that is the
radius has to be of the order of the wavelength). This means that, when the
beam from an extended source is focused inside a gravitational channel at a
distance $L$, an inverted image of the source is formed, having the very
same geometrical dimensions of the source. The channel ''draws'' the source
closer to the observer since, if the true distance of the observer from the
source is $R$, its image brightness will correspond to that of a similar
source at the closer distance

\begin{equation}
R_{eff}=R-l-L\,.
\end{equation}

In the opposite limit, $\omega l\gg 1$, we have $\tan \omega
L\rightarrow \infty $, so that $L\simeq \pi /\omega $, that is the
shortest focal length of the gravitational channel is

\begin{equation}
L_{foc}=\sqrt{\frac{\pi c^{2}}{4G\rho }}\,.
\end{equation}

It is relevant to note that the expression \ $L_{foc}$\ leads to \
values of about 100 Mpc for a typical galactic density $\rho
_{0}\sim 10^{-24}\,g/cm^{3}$. In other words, this can be assumed
as a typical length for large scale structures.

\section{Are there candidates for cosmological refraction channels?}

Early structures, resulting from inflationary phase transitions, could be
good candidates to implement the above filamentary or planar structures
acting as cosmological refraction channels. Essentially, if the primordial
universe undergoes a first or second order phase transition, such a
transition takes place on a short time scale ( $\tau <H^{-1}$ ), and it will
lead to correlation regions inside of which the value of a certain scalar
field $\varphi $ (the inflaton which leads dynamics) is approximately
constant, but outside $\varphi $ ranges randomly over the vacuum manifold.
In the simplest case, for a Ginzburg-Landau-like potential of the form

\begin{equation}
V\left( \varphi \right) =\frac{\lambda }{4}\left( \varphi ^{2}-\varphi
_{0}^{2}\right) ^{2},
\end{equation}

the correlation regions are separated by domain walls, regions in space
where $\varphi $ leaves the vacuum manifold $\mathcal{M}$ and where,
therefore, potential energy is localized. This energy density can act as a
seed for structure.

There are various types of local and global topological defects
(Kibble 1976) (regions of trapped energy density) depending on the
number of real components of $\varphi $ (being, in general,
$\varphi
^{2}=\sum\nolimits_{i=1}^{n}\varphi _{i}^{2}$). In the case of domain walls $%
n=1$.

The rigorous mathematical definition refers to the homotopy of $%
\mathcal{M}$ . The words ''local'' and ''global'' refer to whether
the symmetry, which is broken, is a gauge or global symmetry. In
the case of local symmetries, the topological defects have a well
defined core outside of which $\varphi $ contains no energy
density in spite of non vanishing
gradients $\nabla \varphi $ ; the gauge field can absorb the gradient, i.e. $%
D_{\mu }\varphi =0$ \ when $\partial _{\mu }\varphi \neq 0$. Global
topological defects, however, have long range density fields and forces.

Practically, the topological defects can be classified by their topological
characteristic which, in term of the scalar field $\varphi $ , depends on
the number of real components.

Pointlike defects ($n=3$) give monopoles, linear defects ($n=2$) give cosmic
strings, surfaces-like defects ($n=1$) are domain walls, hypersurface-like
defects, but not properly topological defects, with $n=4$, are textures.

In principle, all these structures can be seeds for large scale structures
depending on the cosmological model we adopt. Taking into account realistic
scenarios, the best candidates seem to be cosmic strings.

Furthermore, any first-order phase transition proceeds via bubble nucleation
so that invoking filamentary and planar structures for channeling effects is
reasonable (local surface portions of large bubbles can be approximated to
planar structures).

For example, cosmic strings easily realize the condition to obtain
a Newtonian potential useful, as discussed above, to produce a
gravitational refraction channel. In fact, in the weak energy
limit approximation, these topological defects are described
internally by the Poisson equation $\nabla ^{2}\Phi =\rho _{0}$
and externally by $\nabla ^{2}\Phi =0$ , where $\rho _{0}$\ is a
constant. The condition on the density gradient is naturally
recovered. Another useful feature is that they act as
gravitational lenses after the quasar formation (Vilenkin 1985;
Gott III 1985) Immediately we get, internally, $\Phi \sim r^{2}\
$and, by the dynamical evolution of the universe, strings can
evolve into structures with lengths of the order $\ \sim 100$ Mpc.
However, also after evolution, strings remain ''wires'' and then,
in order to get the cylindrical structures which we need for
gravitational channels (e.g. a filament of galaxies), we have to
invoke further processes where strings are seeds to cluster matter
at large scales.

Furthermore, the motion of cosmic strings with respect to the
background produces wakes and filaments which are able to evolve
into large scale structures like clusters or filaments of galaxies
(Vachaspati \& Vilenkin 1991).

All these arguments, as it is well known, are hypotheses for large scale
structure formation.

Our issue is now to construct reasonable toy models which, by channeling
effects, could explain, in a simple way, the huge luminosity of quasars and
twin effects in gravitational lensing.

\section{The model}

The major aim of our simulation is to perform a test about the possibility
that twin or brother quasars could be explained by trapped and guided light,
which was emitted by quasar sources and guided thanks to structures acting
like channels. Moreover, the single quasar itself could be a typical object
(a bulge of a standard galaxy) which appears so magnified thanks to the
presence of channelling effects.

In our simulation, as discussed above, we can take into account a
gravitational first order phase transition from which false vacuum
bubbles nucleate inside a true vacuum cosmological background. Our
fundamental assumption is that they cosmologically evolve giving
rise to a cellular structure capable of yielding the observed
large scale structures (Kolb \& Turner 1990). After, we take into
account also filamentary large scale structures.

It is also important to note that, due to the cosmological expansion, the
bubbles continue to grow even after they collide; this is the difference
with models used in crystal growth such as the Voronoi and the Johnson-Mehl
models, where crystals stop to grow in the direction of the neighbour one.

By simple geometric arguments, we can put around a circle other six circles
in a given area, with the same radius. If these circles are expanding, to
minimize the space among them, they degenerate into hexagons. The Fig. 1
shows the situation.

\begin{figure}
\resizebox{\hsize}{!}{\includegraphics{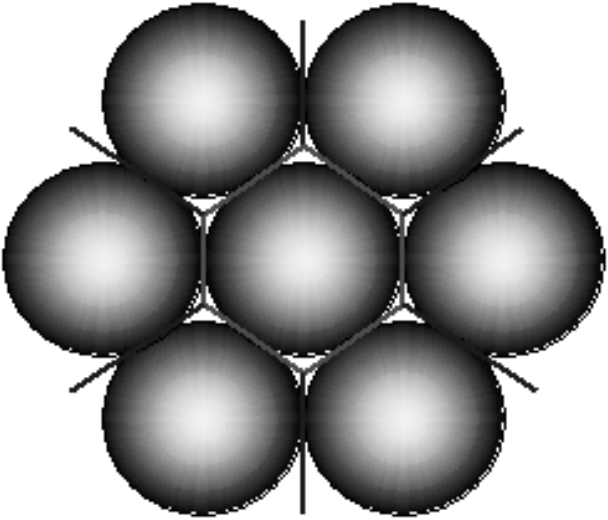}} \hfill
\caption{Model for the formation of contacting planar structures.}
\label{Figure1}
\end{figure}

In other words, density gradients of matter will be generated among the
hexagons in the boundary regions. In three dimension, the spheres will
degenerate into dodecahedrons by forming a typical honeycomb structure with
2-dimensional elementary pentagonal cells (see Fig. 2), whose surfaces will
give rise to planar structures with matter density different from zero.

Finally, given an expanding universe several dodecahedrons, remnants of a
primordial gravitational phase transition, could yield a sort of honeycomb
structure. The isotropy and homogeneity would be preserved at larger scales
(size $\sim 1000$ Mpc).

In this case, the formation of clustered galactic structures will be
naturally due to the contact among several spheres, which would give rise to
density gradients.

\begin{figure}
\resizebox{\hsize}{!}{\includegraphics{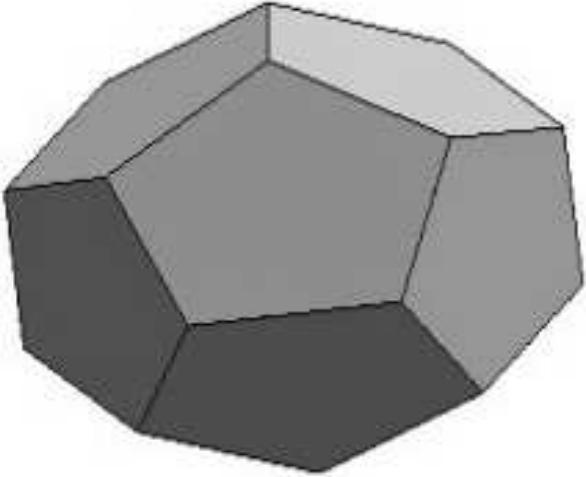}} \hfill
\caption{Elementary cell like dodecaedron.} \label{Figure2}
\end{figure}

Such planar structures on contact surfaces could act as refraction
channels, e.g. they could trap and guide the light due to the
quasi-constant surface density; in this way, we should see a
universe like that observed. To be more precise, we do not see an
exact honeycomb, but a selfsimilar structure (Sylos Labini,
Mountuori \& Pietronero 1998) like those described in fractal
models: in a more realistic model, we have to consider several
spheres with different radii $R_{i}$ (we have not a perfect
dodecahedron, but a more complicate polyhedron, see for example,
Fig. 3). Dynamics will be governed by the symmetry breaking during
the phase transition: vacuum density could be different in
different bubbles. However, we can obtain relevant results just
taking into account balls with the same radius, that is, by
considering bubble nucleation at same density and expanding rate
at constant speed $v$. Similar considerations can be done also for
filamentary structures which act as refraction channels.

\begin{figure}
\resizebox{\hsize}{!}{\includegraphics{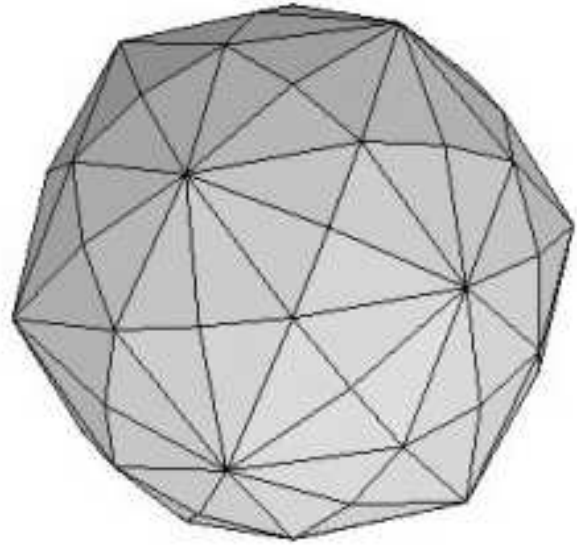}} \hfill
\caption{Elementary cell like trapezoidal icositetrahedron.}
\label{Figure3}
\end{figure}

\subsection{Hypotheses and previsions of the simulation}

Let us consider a universe model with an observed radius
$R_{0}=3000$ Mpc. Furthermore, let us assume $10^4$ quasars
uniformly and isotropically distributed in this spherical space.
The positions of the objects are fixed in a random way by using a
random number generator. Moreover, the quasars are point-like
objects. Furthermore, it is supposed that every single quasar
emits light in an isotropic way.

Between the refraction channels, we have to distinguish filamentary
structures and planar structures.

We produce randomly distributed filamentary structures with a dimension of
about 100 Mpc along the major axis and a shape like a cylinder with 5
points, where the longitudinal axis of the cylinder changes its direction. A
typical simulated channel is shown in the Fig. 4.

In particular, the channels of the model are built by five
cylinders with a length of $ 1\div 20$ Mpc and radius $r \sim
1\div 100$ Kpc. The positions of the channel starting points, that
is the centres of the first base circle in the first cylinder and
the longitudinal axis of the cylinder are randomly generated.

\begin{figure*}[tbp]
\resizebox{12 cm}{!}{\includegraphics{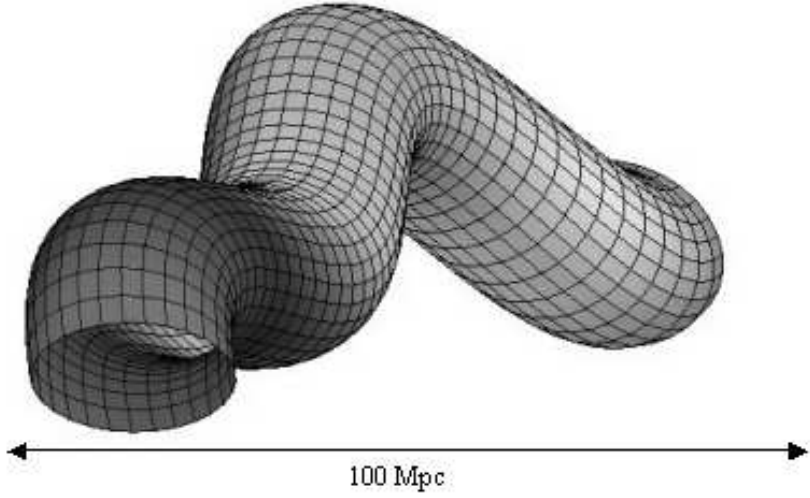}} \hfill
\parbox[b]{55 Mm}{
\caption{A typical simulated channel.} \label{Figure4}}
\end{figure*}

There is a constraint on this generation: we assume that the
cylinders cannot collapse in a compact object. In other words, by
defining $\theta_i$ and $\phi_i$ the angles in the links between
two consecutive cylinders along their longitudinal axis (see Fig.
5), we impose:

\begin{equation*}
0<\vartheta _{i}<\pi /2\qquad and\qquad 0<\varphi _{i}<2\pi ,
\end{equation*}

\begin{figure*}[tbp]
\resizebox{12 cm}{!}{\includegraphics{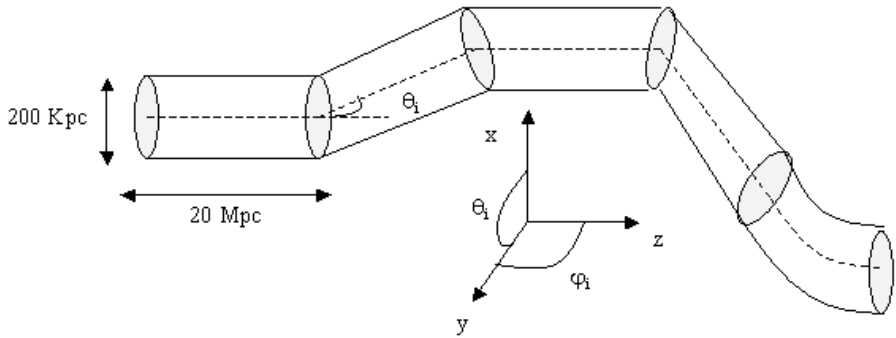}} \hfill
\parbox[b]{55 mm}{
\caption{Scheme of a channel.} \label{Figure5}}
\end{figure*}

with $\vartheta _{i}$ and $\varphi _{i}$ randomly selected.

Alternatively, we can have casually distributed planar structures acting
like refraction channels, with linear dimensions of about $100$ Mpc, on
pentagonal surfaces.

In particular, in a universe with a radius $R_{0}=3000$ Mpc, we
can inscribe several pseudo-balls $b_i$ with a radius $R_{i}$ such
that the linear dimensions have an extension of about $l_{i}=100$
Mpc. This can be made in different ways:

1) Let us consider a circular sector with an angular spread of $\pi /24=7.5%
%TCIMACRO{\UNICODE[m]{0xb0}}%
%BeginExpansion
{{}^\circ}%
%EndExpansion
$; by assuming that the linear extension of the structure is of
100 Mpc, we obtain:

\begin{equation}
100\ \text{Mpc}=l_{i}=\frac{\pi }{24}R_{i}\quad \rightarrow \quad R_{i}=764\
\text{Mpc.}
\end{equation}

In this case, we can inscribe $n_{balls}=60$ balls with a radius $R_{i}$ in
the universe with a radius $R_{0}$. Moreover, we have to calculate the error
by assuming

\begin{equation*}
2\pi R_{i}=4800\ \text{Mpc},\ l=100\ \text{Mpc}\quad \rightarrow
\quad \epsilon =\frac{l}{2\pi R_{i}}=2\%.
\end{equation*}

2) By assuming a circular sector with an angular spread of $\pi /18=10%
%TCIMACRO{\UNICODE[m]{0xb0}}%
%BeginExpansion
{{}^\circ}%
%EndExpansion
$, we obtain

\begin{equation*}
R_{i}=573\ \text{Mpc},\quad n_{balls}=143,\quad \epsilon =3\%.
\end{equation*}

3) By assuming a circular sector with an angular spread of $\pi /12=15%
%TCIMACRO{\UNICODE[m]{0xb0}}%
%BeginExpansion
{{}^\circ}%
%EndExpansion
$, we obtain

\begin{equation*}
R_{i}=382\ \text{Mpc},\quad n_{balls}=484,\quad \epsilon =4\%.
\end{equation*}

4) By assuming a circular sector with an angular spread of $\pi /10=18%
%TCIMACRO{\UNICODE[m]{0xb0}}%
%BeginExpansion
{{}^\circ}%
%EndExpansion
$, we obtain

\begin{equation*}
R_{i}=318\ \text{Mpc},\quad n_{balls}=840,\quad \epsilon =5\%.
\end{equation*}

5) Finally assuming a circular sector with an angular spread of $\pi /9=20%
%TCIMACRO{\UNICODE[m]{0xb0}}%
%BeginExpansion
{{}^\circ}%
%EndExpansion
$, we obtain

\begin{equation*}
R_{i}=286\ \text{Mpc},\quad n_{balls}=1154,\quad \epsilon =6\%.
\end{equation*}

In the simulation, it is useful to implement an attenuation mechanism to
make the model more realistic. If the sources are at distances of the order
of $1000$ Mpc, since the luminosity decreases as $R^{-2},$ we assume that
only fluxes large enough to survive to this physical cut can be detected as
the channelling effects intervene. Otherwise they are completely absorbed by
intergalactic media.

Next step is to verify how much light is trapped by the intervening channels
and how many sources are lensed and then appear as twin images to the
observer.

\subsection{The implementation}

The simulation is made by using the \textit{New Object Visual
Programming Technique} (National Instruments: Reference Manual
1998; National Instruments: Functions Manual 1998; National
Instruments: G-Math Manual 1998).

The simulation is built up by a main program, that calls and manages a
collection of subprocesses described as follows.

1) The first process is built up by 5 secondary sequences:

\qquad 1.1) The first one - sequence 1.1 - gives the quasar
coordinates which are randomly generate;

\qquad 1.2) The second one - sequence 1.2 - gives the photons
coming out from quasars: in this step the arrival points of coming
out photons are evaluated;

\qquad 1.3) The third one - sequence 1.3 - generates the light rays at each
quasar position;

\qquad 1.4) The fourth one - sequence 1.4 - rejects all photons out of $%
2R_{0}=6000$ Mpc (the boundary of our toy universe);

\qquad 1.5) The fifth one - sequence 1.5 - prepares a multidimensional array
and a file to perform and to save the intersections between the rays and the
waveguides;

2) The second process regards the objects, whose light is picked up, trapped
and guided by the structures (it gives the twined quasars);

3) The third process has two sequences:

\qquad 3.1) generates the refraction channel in random positions into the
spherical volume representing the universe;

\qquad 3.2) evaluates the parameters to intersect the rays from the quasars
(this subprocess is completely equivalent to 1.5);

4) The fourth process evaluates the attenuation function for the light
coming from quasars;

5) The fifth process prepares the trapping;

6) The sixth process performs the trapping and guiding of light in the
channels;

7) The seventh process tests that the same light rays is not
counted more than one time, if it intersects more sub-cylinders of
the same channel.

In Fig. 6, there are shown several examples of simulated
refraction channels.

\begin{figure*}
\resizebox{12 cm}{!}{\includegraphics{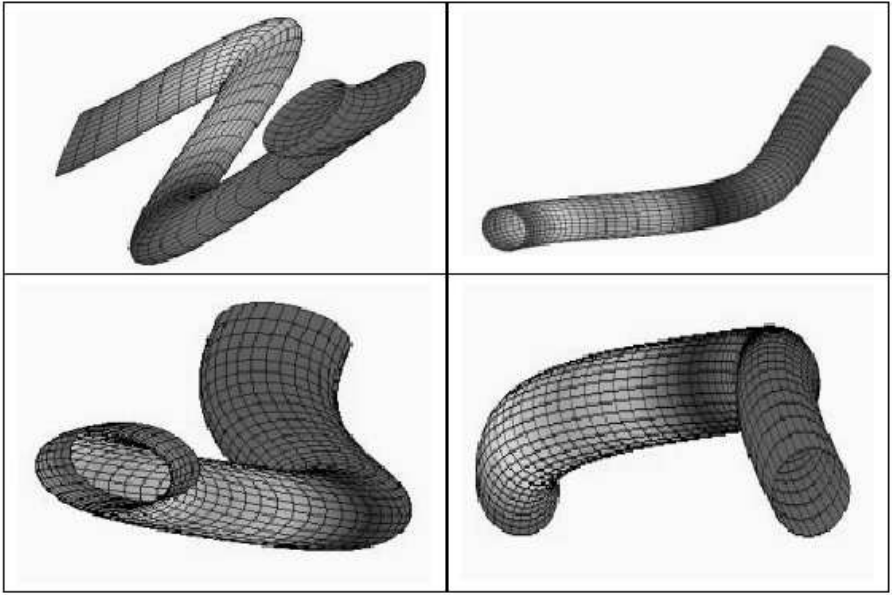}} \hfill
\parbox[b]{55 mm}{
\caption{Some examples of simulated refraction channels.}
\label{Figure6}}
\end{figure*}

\section{The results}

The results of the simulation are summarized, by considering: 1) a
fixed number of channels chosen in relation to $R_i$ for the
planar case and 200 for the filamentary one; 2) $10^4$ quasars
randomly fixed as input.

a) The quasar positions $R$ (in Mpc) and the redshifts $Z,$ after
the attenuation, have a Gaussian distribution fit with the mean
value and the standard deviation in full agreement with
theoretical and observative results (Capozziello \& Iovane 1999;
Fort \& Mellier 1994; Broadhurst, Taylor, Peakock 1995);

\begin{figure*}
\resizebox{12 cm}{!}{\includegraphics{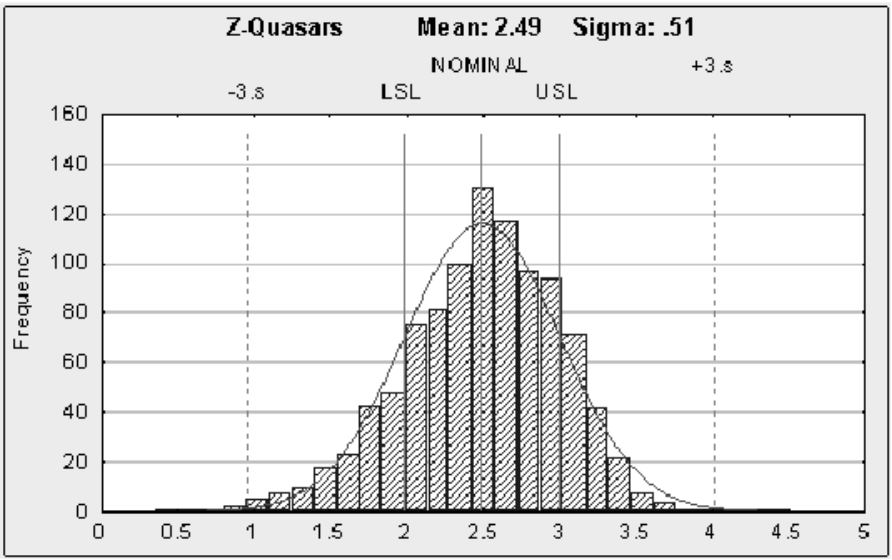}} \hfill
\parbox[b]{55 mm}{
\caption{Quasar redshift $Z$ after attenuation process.}
\label{Figure7}}
\end{figure*}

b) Also for the twined quasars, $R$ and $Z$ are comparable with the
theoretical models; in particular, the twined quasars are a subset of the
input quasars and they are also isotropically distributed in the planes $x-y
$ and $x-z$.

\begin{figure*}[tbp]
\resizebox{12 cm}{!}{\includegraphics{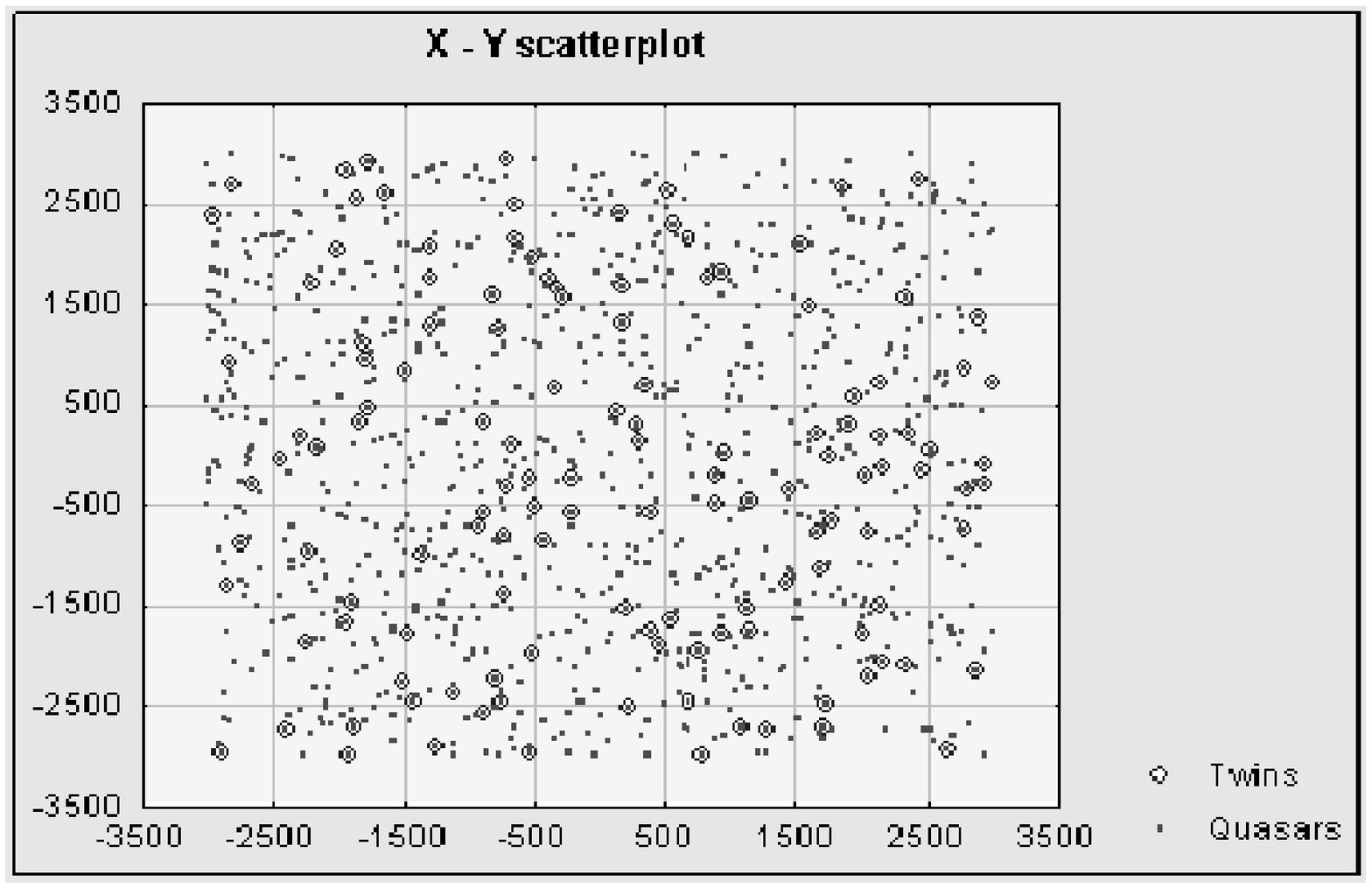}} \hfill
\parbox[b]{55 mm}{
\caption{Scatter plot $x-y$ plane of quasars after attenuation and
twined quasars.} \label{Figure8}}
\end{figure*}

\begin{figure*}[tbp]
\resizebox{12 cm}{!}{\includegraphics{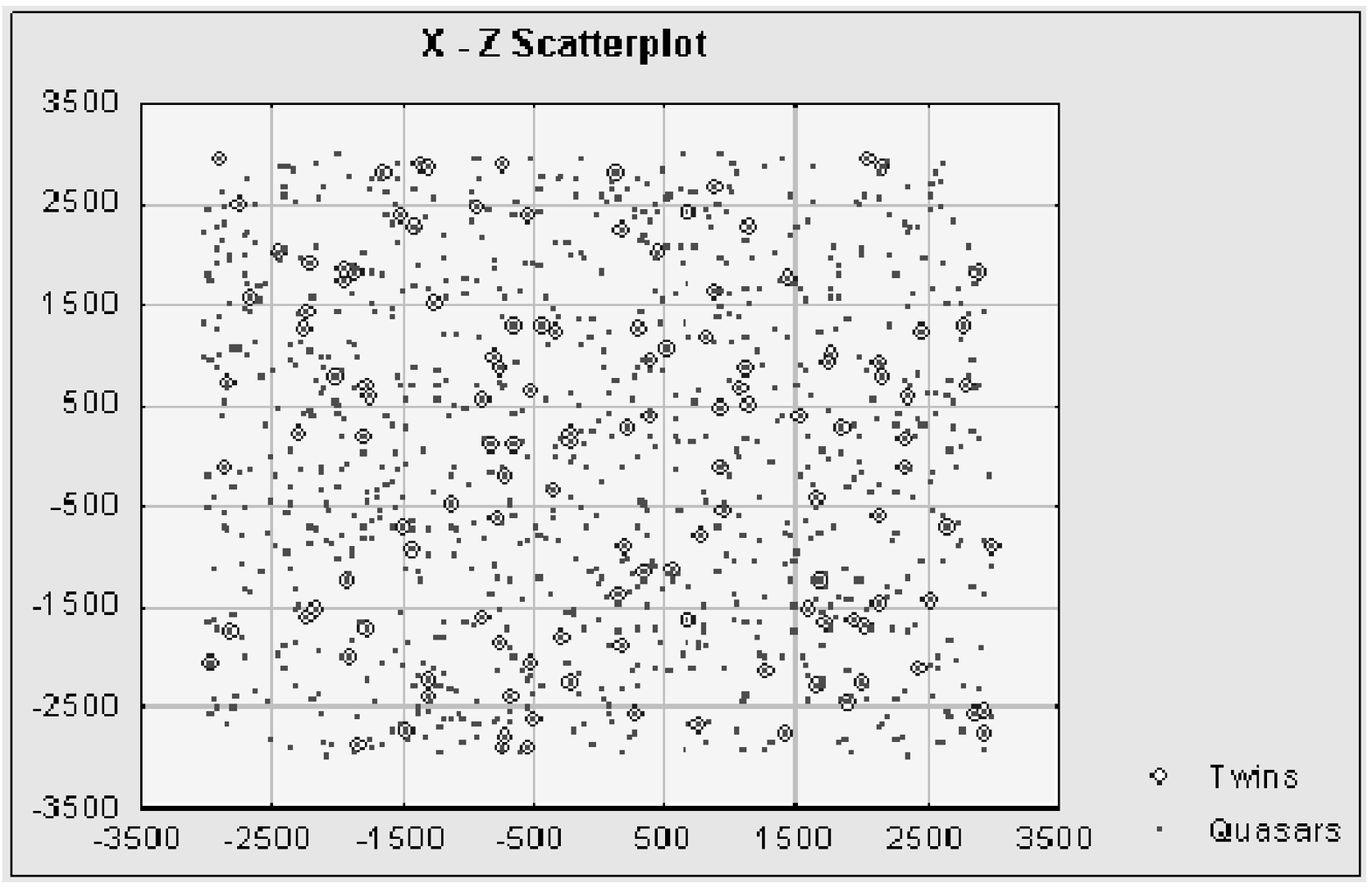}} \hfill
\parbox[b]{55 mm}{
\caption{Scatter plot $x-z$ plane of quasars after attenuation and
twined quasars.} \label{Figure9}}
\end{figure*}

c) The Table 1 summarizes the obtained results for the planar
channels from a numerical point of view.

\begin{table*}
\begin{tabular}{|c|c|c|c|c|}
   \hline
   {\bf Input number of quasars} & {\bf number of channels} & {\bf number of quasars} &
   {\bf Twined quasars} & {\bf Twined Q./Q. after atten. }\\
   \hline
    $10^4$ & 60 & 1015 & 109 & 7\% \\
    \hline
    $10^4$ & 143 & 977 & 141 & 14\% \\
    \hline
    $10^4$ & 484 & 995 & 260 & 26\% \\
    \hline
    $10^4$ & 840 & 1023 & 337 & 33\% \\
    \hline
    $10^4$ & 1154 & 981 & 461 & 35\% \\
  \hline
\end{tabular}
  \caption{Tab. 1. Simulation summary.}
\end{table*}

About 10\% of initial quasars pass trough the attenuation process.
A fraction, from 7\% to 35\% of quasars after the attenuation,
undergoes a channelling effect. This fraction is linked to the
number of channels. The number of twined quasars increases as the
number of channels increases, by fixing the number of quasars
after the attenuation. A preliminary fit can be made between the
number of refraction channels and the twined quasars. Fig. 10
shows a linear correlation.

\begin{figure*}[tbp]
\resizebox{12 cm}{!}{\includegraphics{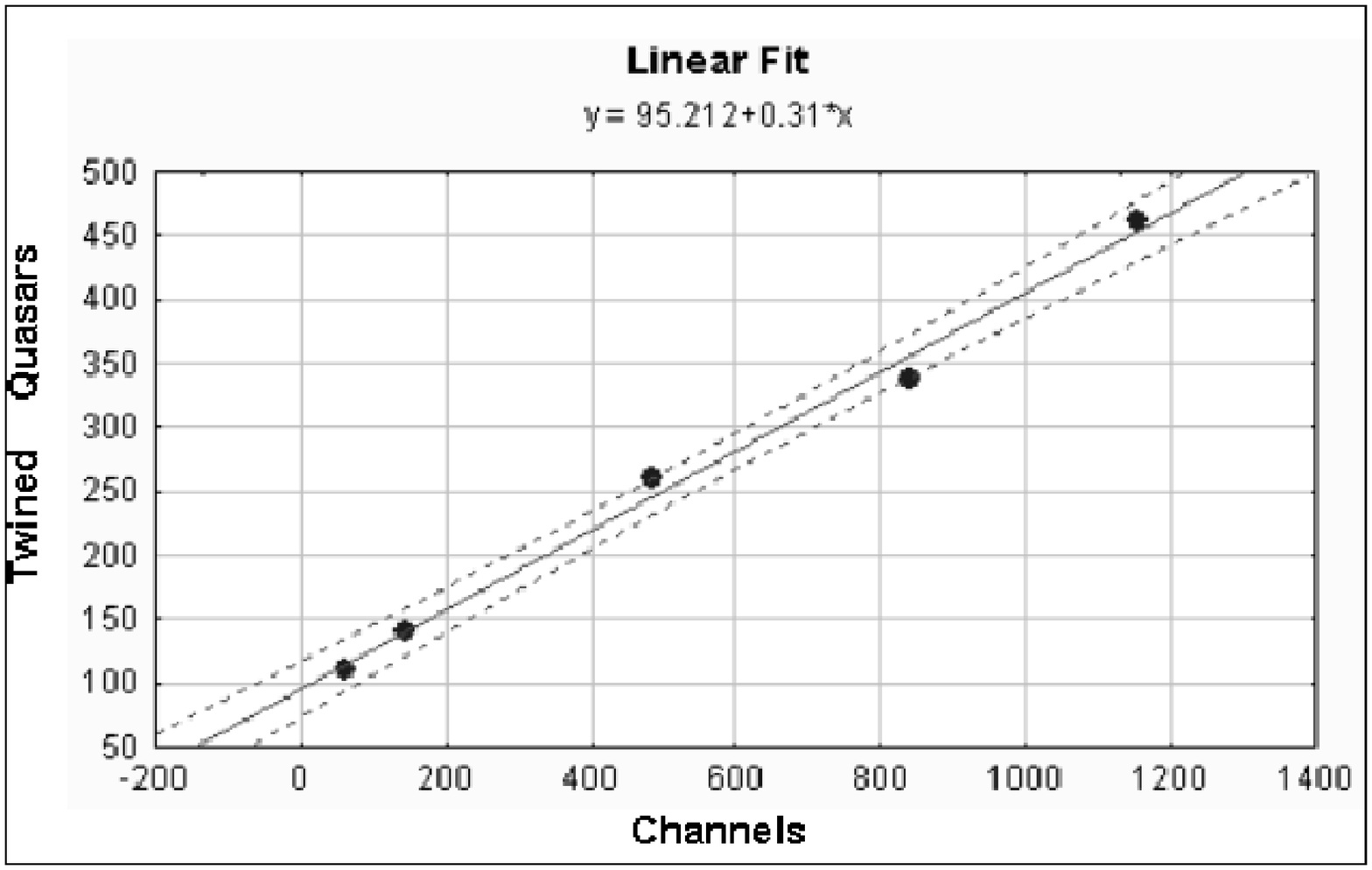}} \hfill
\parbox[b]{55 mm}{
\caption{Linear fit between the number of channels and twined
quasars.} \label{Figure11}}
\end{figure*}

The results are in full agreement with the filamentary channels. In fact, in
this case, we have 982 quasars after the attenuation and, among them, 137
are twined. In other words, about 10\% of the initial quasars pass trough
the attenuation process and a fraction of about 14\% of these ones undergoes
a channeling effect.

In a first approximation, a superposition of the two distributions
( not twined quasars after attenuation and twined quasars) leads
our analysis in the same direction of the raw theoretical model,
that is, we get a Poisson-like distribution for all quasars. On
the other hand, a finest analysis shows two peaks in the quasars
distribution: the highest is due to non-twined quasars, whose
light is not trapped and guided in the channels and the shortest
is due to twined quasars . This fraction is in the range 7\%-35\%.

\begin{figure*}[tbp]
\resizebox{12 cm}{!}{\includegraphics{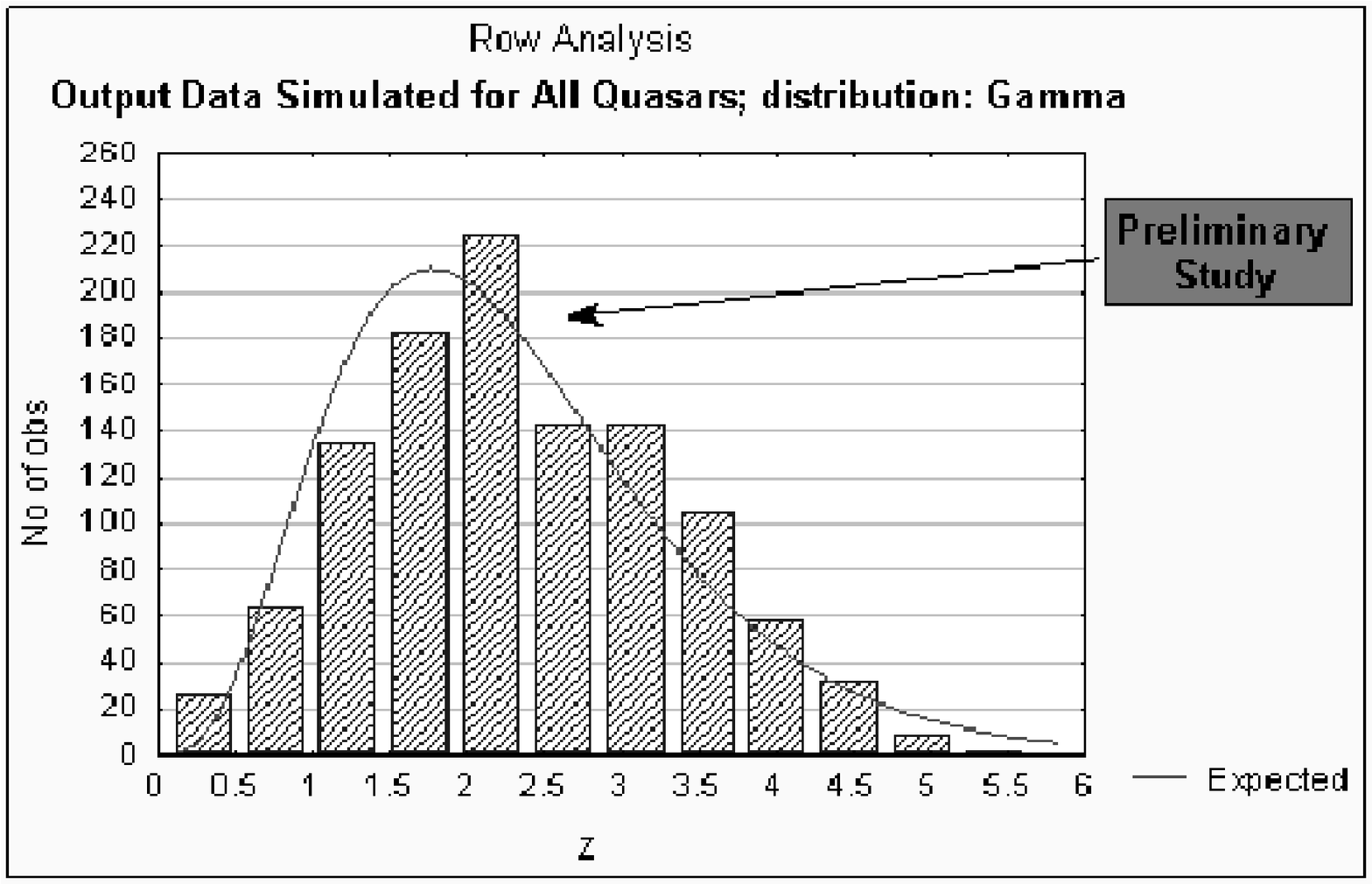}} \hfill
\parbox[b]{55 mm}{
\caption{Expected Poisson shape in $Z$ quasars distribution like
in raw theoretical models.} \label{Figure12}}
\end{figure*}

\begin{figure*}[tbp]
\resizebox{12 cm}{!}{\includegraphics{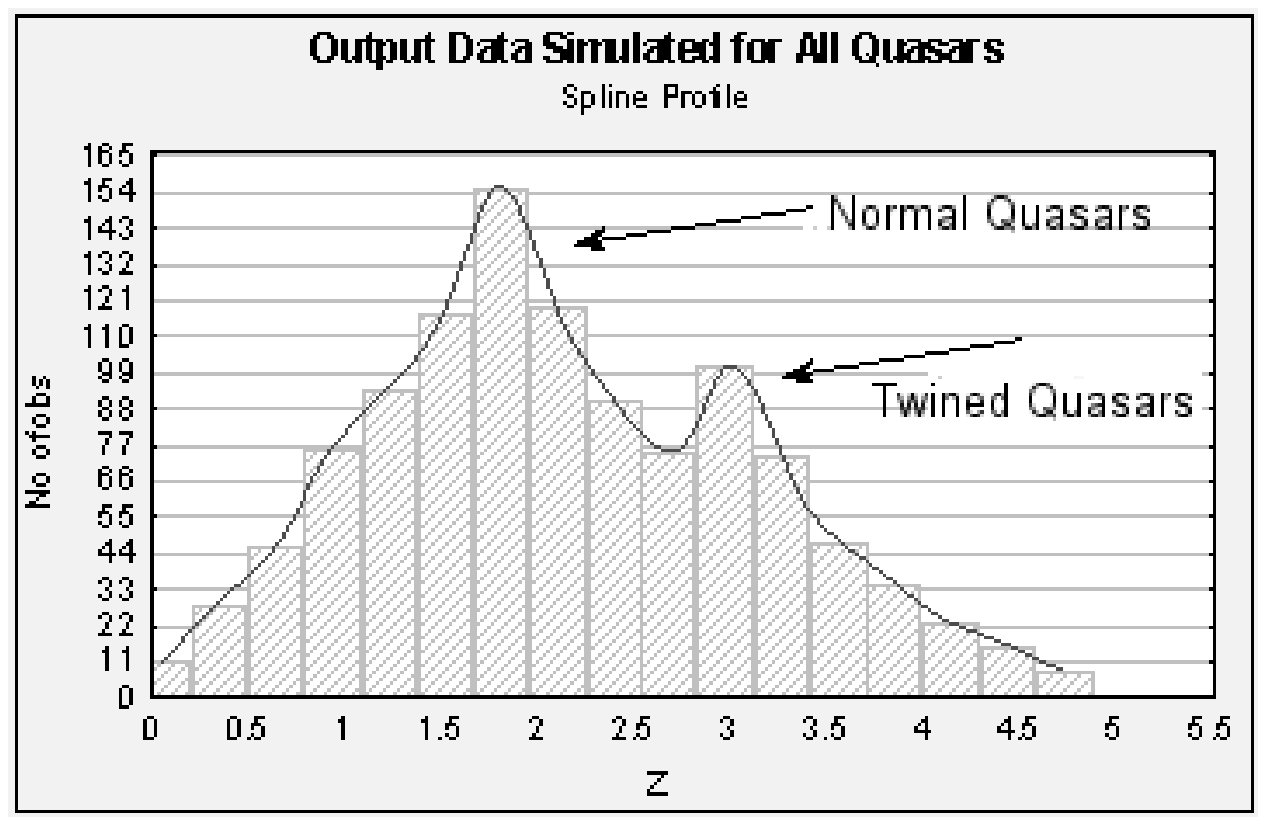}} \hfill
\parbox[b]{55 mm}{
\caption{Double peak in quasar distribution like in observations
(CRONARIO 1998).} \label{Figure13}}
\end{figure*}

\section{Discussion and conclusion}

\bigskip In this paper, we propose a mechanism by which it is possible to
explain two of major puzzles of quasar theory: the huge luminosity
and the twin (or brother) objects. Both of them, till now, have
invoked exotic explanations as the presence of a supermassive
black hole inside the core radius or lensing effects, where the
position and the size of intervening deflector (usually a galaxy)
has to be in same sense ''fine tuned''. By our cosmological
refraction channeling effects both the issues are addressed in a
natural way.

The huge luminosity at far red-shift is nothing else but the
effect of the channel which ''draws'' the luminosity of the source
closer to the observer. In this sense, quasars could be just
primordial (in same sense ''ordinary'') galaxies whose luminosity
in not dispersed as $1/R^{2}$ This fact means that the source
brightness will turn out to be much stronger than the brightness
of analogous objects located at the same distance (i.e. at the
same redshift \textit{Z}) and the apparent energy released by the
source seems anomalously large. Besides, the large angular
distance separation of twins and brothers is due to the geometry
of (filamentary or planar) channels without posing ''ad hoc'' lens
galaxies between the source and the observer. This hypothesis, as
shown, naturally explain the double peak luminosity function of
quasars.

However, what we have discussed here is nothing else but a toy
model. In fact, the discrete nature of matter distribution within
the channels affect the light propagation (we have to take into
account scattering points, dissipation, absorbtion, and dispersion
in a more realistic model).

Depending on the distribution of discrete matter, we could have several
places where the light could lake out.

Furthermore, the distribution of our filamentary and planar structures has
to be compared to the microwave background in order to preserve the overall
homogeneity and isotropy of observed very large scale structure.

All these issues deserve further studies and the comparison to the
observational data.

\begin{acknowledgements}
The authors wish to thank J.Wampler for the suggestions and comments which
allow to improve the paper.
Work supported by fund ex 60\% D.P.R. 382/80.
\end{acknowledgements}

\end{document}